\documentclass[a4paper]{iopart}

\usepackage{graphicx}
\begin{document}

\paper{Electronic structure of the noncentrosymmetric superconductor Mg$_{10}$Ir$_{19}$B$_{16}$}

\author{B Wiendlocha, J Tobola and S Kaprzyk}

\address{Faculty of Physics and Applied Computer Science, AGH University of Science and Technology, Al. Mickiewicza 30, 30-059 Cracow, Poland}
\ead{bartekw@fatcat.ftj.agh.edu.pl}
\begin{abstract}
Electronic structure of a novel superconducting noncentrosymmetric compound Mg$_{10}$Ir$_{19}$B$_{16}$ was calculated using the Korringa-Kohn-Rostoker method. Electronic part of the electron-phonon coupling constant, McMillan-Hopfield parameters, were calculated using the rigid-muffin-tin approximation (RMTA). The magnitude of the electron-phonon coupling constant $\lambda$, analysing atomic contributions, is discussed. Our results show, that superconductivity in Mg$_{10}$Ir$_{19}$B$_{16}$ is presumably mediated by electron-phonon interaction.

\end{abstract}



\section{Introduction}
The interest in noncentrosymmetric structures, exhibiting superconductivity, has grown up in the past years. There are only a few examples of this type, which belong to various classes of materials,
e.g. an antiferromagnetic heavy fermion system CePt$_3$Si~\cite{cept3si}, ferromagnetic uranium compound UIr, superconducting under pressure~\cite{uir}, or non-magnetic ternary borides Li$_2$Pd$_3$B~\cite{lipdb1} and Li$_2$Pt$_3$B~\cite{liptb}. The main reason, why these systems are especially attracting, is related to the role of inversion symmetry in electron pairing. The absence of inversion symmetry may suppress the triplet pairing or mix singlet and triplet symmetry \cite{anderson,sym1,sym2}.

Very recently, Klimczuk and co-workers \cite{klimczuk1} synthesised a new type of intermetallic light-element based compound Mg$_{10}$Ir$_{19}$B$_{16}$, exhibiting superconductivity near 5~K. This novel material also belongs to the rare noncentrosymmetric structures, and crystallises in a large and rather complex $bcc$ cell (space group I-43$m$)~\cite{klimczuk1}.
Mg$_{10}$Ir$_{19}$B$_{16}$ is partly similar to the Li$_2$(Pd,Pt)$_3$B system, since these structures contain an alkali metal (Li, Mg), boron, and heavy transition metal (Pd, Pt, Ir).

In this work we intend to start the discussion on superconductivity mechanism in this unusual compound, analysing the electronic structure and the strength of electron-phonon coupling (EPC). Assuming the BCS-type behaviour in Mg$_{10}$Ir$_{19}$B$_{16}$, we study whether superconductivity is driven by light boron sublattices, like e.g. in MgB$_2$, or by heavy transition metal atoms, as suggested for Li$_2$Pd$_3$B~\cite{lipdb-pickett}.

\subsection{Computational details}

Electronic structure calculations were performed using the Korringa-Kohn-Rostoker (KKR) multiple scattering method \cite{prb99, jpcm04}. The crystal potential was constructed in the framework of the local density approximation (LDA), using von Barth and Hedin formula \cite{lda} for the exchange-correlation part. For all atoms angular momentum cut-off $l_{max}=3$ was set, \makebox{{\bf k}-point} mesh in the irreducible part of Brillouin zone (IRBZ) contained over 200 points. Densities of states (DOS) were computed using the tetrahedron \makebox{{\bf k}-space} integration technique, generating over 600 tetrahedrons in IRBZ. Due to the high atomic number of iridium ($Z=77$) semi-relativistic calculations were performed, but neglecting the spin-orbit coupling, which is commented at the end of this paper.

\begin{figure}[htb]
\includegraphics[scale=0.65]{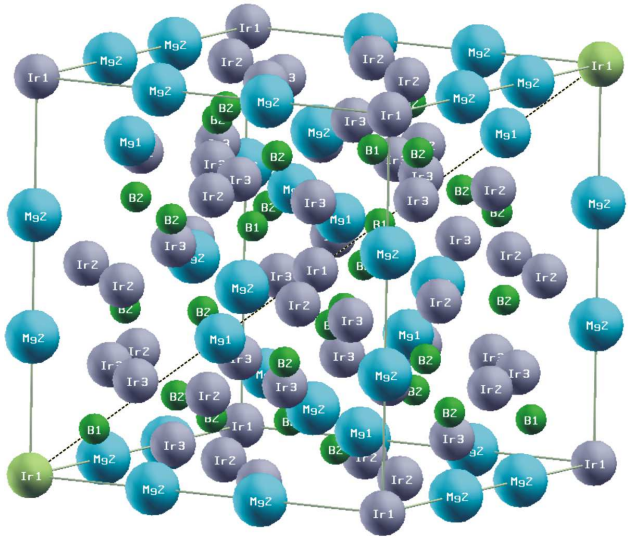}
\caption{Unit cell of Mg$_{10}$Ir$_{19}$B$_{16}$. The lack of inversion centre is clearly seen e.g. along the main diagonal, where B1 and Mg1 atoms break the inversion symmetry. Generated by {\tt XCRYSDEN} \cite{xcrys}.}\label{fig:struct}
\end{figure}

As far as the crystal structure is concerned, experimental lattice constant \makebox{$a = 10.568$~\AA} and atomic positions \cite{klimczuk1} were accounted for the computation (for clarity also shown in \tref{tab:dist}).
Atoms in the unit cell were surrounded by muffin-tin (MT) spheres with following radii: \makebox{$R_{Mg}= 2.82$}, \makebox{$R_{Ir}=2.50$}, \makebox{$R_{B}=1.40$} (in atomic units), filling about 60\% of the cell volume.
In the primitive cell of this system, 45 atoms occupy 7 inequivalent sites, which all are listed in \tref{tab:dist}.
The noncentrosymmetricity of this system is important, and cannot be regarded as the effect of a lattice distortion, as observed e.g. in UIr~\cite{lipdb-pickett}.
The inversion symmetry is broken by both boron sublattices (B1 and B2), as well as Mg1 and Ir3 sites, thus the crystal has a half of symmetry operations of the cubic group O$_h$, i.e. only 24 operations. Among of all sublattices, the positions of iridium atoms are the closest to have inversion symmetry. Since Ir3 occupies $(x,z,z)$ sites, with $x\simeq0.07$, the full cubic symmetry is restored after shifting this position to $(0,z,z)$.

The analysis of superconducting properties is based on the computed McMillan-Hopfield (MH) parameters $\eta$ \cite{mcm,hop} which determine electronic part of the electron-phonon interaction, and directly enter the formula for the electron-phonon coupling constant $\lambda$:

\begin{equation}\label{eq:lambda}
\lambda = \sum_i \frac{\eta_i}{M_i\langle \omega_i^2 \rangle}.
\end{equation}
In equation \eref{eq:lambda}, $\eta_i$ is the MH parameter for each nonequivalent atom $i$ which is characterised by the atomic mass $M_i$ and averaged squared vibrational frequency $\langle \omega_i^2 \rangle$.
This equation divides the electron-phonon coupling constant into site-dependent parts, thus it allows to justify which sublattice gives the most important contribution to the total $\lambda$.
Calculations of MH parameters were preformed
using the rigid muffin tin approximation (RMTA) \cite{gasgy,gomgy,papa,pickett}, which gives the following expression for $\eta$ at each atomic site $i$:
\begin{equation}\label{eq:eta}
\fl
\eta_i  =  \sum_l \frac{(2l + 2)n^i_l(E_F)n^i_{l+1}(E_F)}{(2l+1)(2l+3)N(E_F)}
  \left|\int_0^{R^i_{MT}}dr\ r^2 R^i_l(r)\frac{dV_i(r)}{dr}R^i_{l+1}(r) \right|^2.
\end{equation}
Here, $l$ is the angular momentum number, $n^i_l(E_F)$ is the partial (angular-decomposed) density of states per spin at the Fermi energy ($E_F$), $R^i_l(r)$ is a regular solution of the radial Schr\"odinger
equation, normalised to unity inside the $MT$ sphere of radius $R^i_{MT}$, and $V_i(r)$ is the self-consistent, spherically-symmetric potential. $N(E_F)$ is the total DOS at $E_F$ per spin and per cell.
For more detailed discussion of approximations involved in this method, see e.g. \cite{sc3xb,pss-b} and references therein.


\section{Results and discussion}

\begin{figure}[t]
\makebox{\includegraphics[scale=0.3]{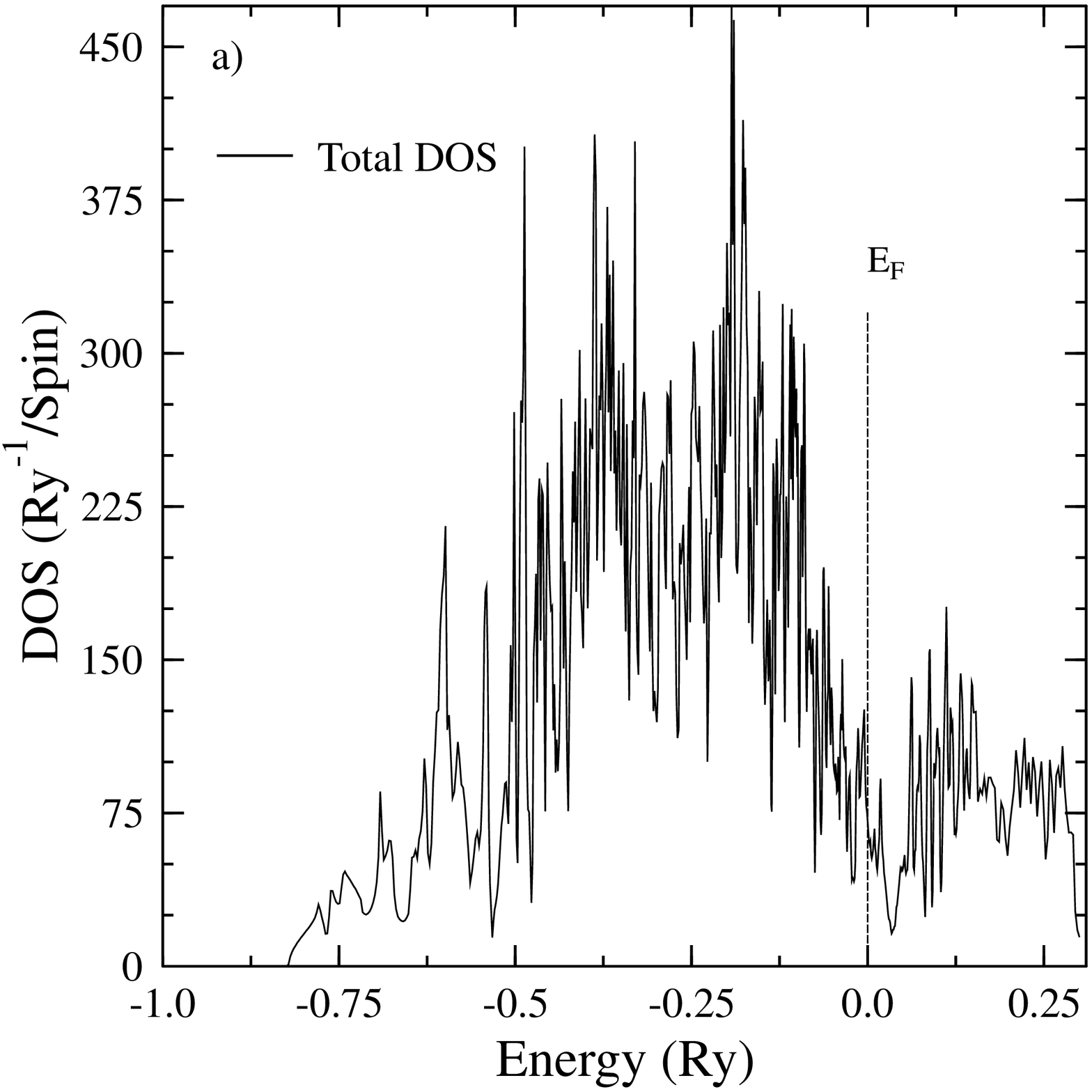}
\includegraphics[scale=0.3]{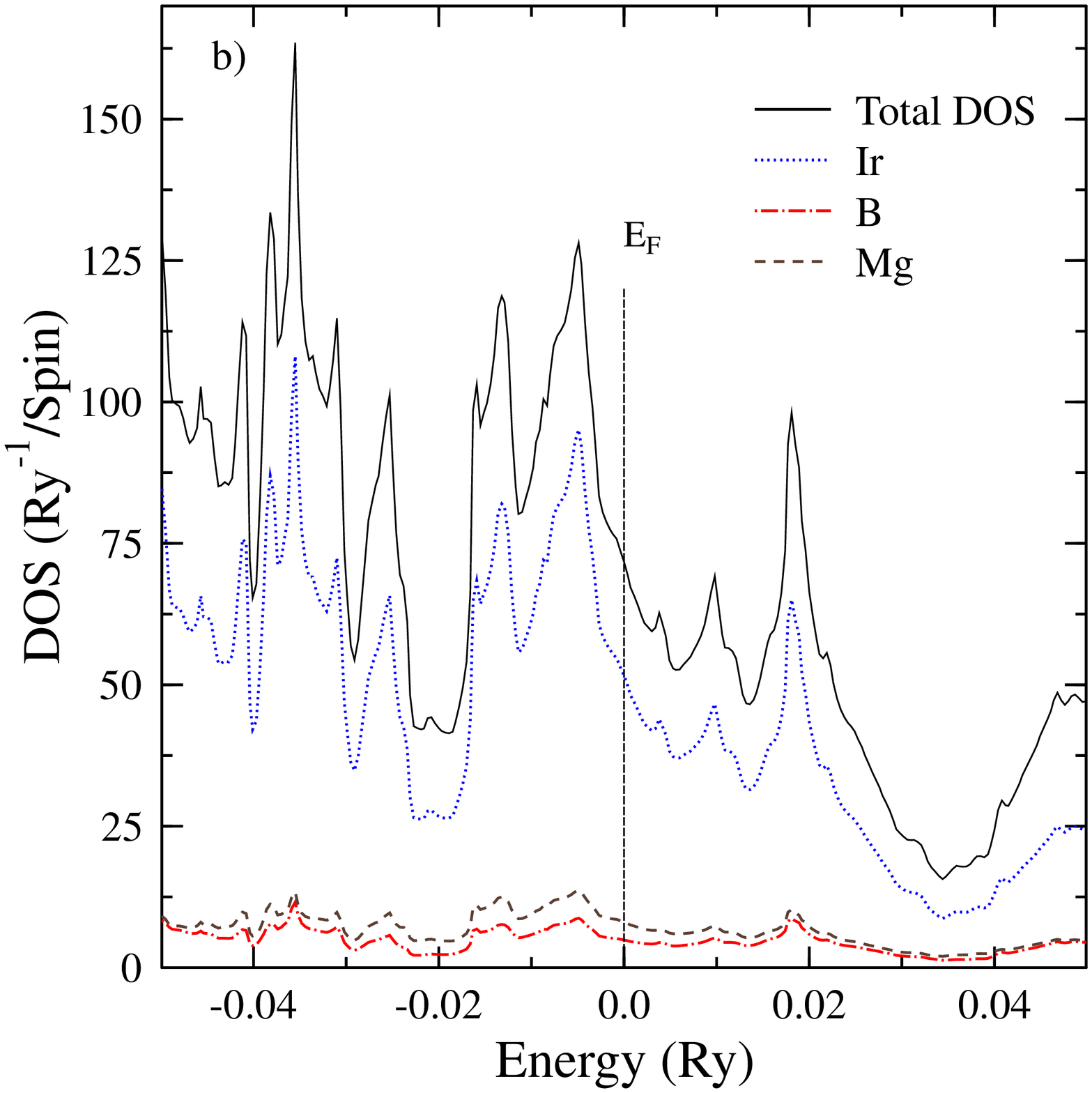}}
\caption{(a) Total DOS of Mg$_{10}$Ir$_{19}$B$_{16}$. (b) Zoom near the Fermi level ($E_F = 0$) with atomic contributions.}\label{fig:dos}
\end{figure}

\begin{figure}[b]
\includegraphics[width=\textwidth]{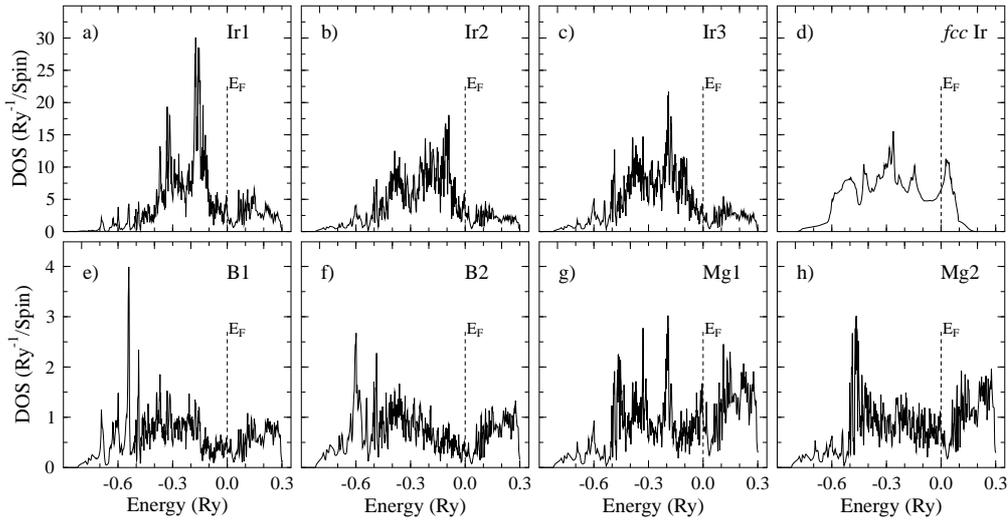}
\caption{Site-decomposed DOS of Mg$_{10}$Ir$_{19}$B$_{16}$ ($E_F=0$). The total DOS of $fcc$ iridium metal is given for comparison in panel (d).}\label{fig:dos2}
\end{figure}

The calculated total and atomic densities of states functions are presented in \fref{fig:dos}.
As one could expect, electronic structure is dominated by iridium $5d$ states. The total DOS curve may be characterised as a collection of numerous van Hove singularities, reflecting the large number of atoms in the unit cell and various interatomic distances.
The site-decomposed DOS are shown in \fref{fig:dos2}, and their values at the Fermi level are gathered in \tref{tab:eta}. The total DOS at Fermi level is about $N(E_F) \simeq 150$ Ry$^{-1}$ per formula unit. The densities at $E_F$, calculated per atom, are rather low (average value per atom: 3.3 Ry$^{-1}$), and generally $E_F$ is located outside of dominating peaks of DOS, on a small decreasing slope of Ir DOS.
Strong hybridisation of $d$ states of Ir and $p$ states of B and Mg is manifested in the separation of bonding and anti-bonding states, and $E_F$ is located in the DOS valley (\fref{fig:dos} and \fref{fig:dos2}).
The interatomic distances, listed in \tref{tab:dist}, supports the enhanced $p$-$d$ hybridisation, especially  between Ir and B atoms (the smallest distances, 2.1 - 2.2 \AA).

\begin{table}[t]
\caption{\label{tab:eta}Densities of states $n(E_F)$, $l$-decomposed DOS $n_l$ (Ry$^{-1}$/spin per atom) McMillan-Hopfield parameters $\eta$, MH parameters for each scattering channel $\eta_{l,l+1}$ (mRy/a.u.$^2$ per atom), and summary MH parameters for each site $w_i\times\eta_i$ (mRy/a.u.$^2$~per~site). $w_i$ is the number of atoms occupying site in primitive cell.}
\lineup
\begin{small}
\begin{tabular}{@{}ll|lllll|lllll}
\br
Atom &  $w_i$ & $n(E_F)$ & $n_s$ & $n_p$ & $n_d$ & $n_f$ & $\eta_i$  & $\eta^{sp}$ &$\eta^{pd}$ & $\eta^{df}$ & $w_i\times\eta_i$\\
\mr
Ir1  & \01 & 2.77 & 0.15 & 0.88 & 1.86 & 0.03 & 2.8& 0.1 & 1.4 & 1.3 &  \02.8\\
Ir2  & \06 & 3.16 & 0.07 & 0.34 & 2.71 & 0.03 & 3.3& 0.0 & 0.8 & 2.5 &  19.8\\
Ir3  & 12  & 2.41 & 0.06 & 0.34 & 1.97 & 0.03 & 2.6& 0.0 & 0.6 & 2.0 &  31.2\\
B1   & \04 & 0.40 & 0.03 & 0.35 & 0.01 & 0.00 & 1.2& 0.0 & 1.2 & 0.0 &  \04.8\\
B2   & 12  & 0.27 & 0.04 & 0.22 & 0.01 & 0.00 & 0.6& 0.0 & 0.6 & 0.0 &  \07.2\\
Mg1  & \04 & 0.98 & 0.17 & 0.69 & 0.10 & 0.03 & 0.7& 0.5 & 0.2 & 0.0 &  \02.8\\
Mg2  & \06 & 0.67 & 0.09 & 0.41 & 0.14 & 0.03 & 0.2& 0.1 & 0.1 & 0.0 &  \01.2\\
\br
\end{tabular}
\end{small}
\end{table}

It is also interesting to compare the computed site-decomposed Ir densities, to the DOS of metallic $fcc$ iridium, which is presented in \fref{fig:dos2}(d). The shape of lower part of DOS in Ir2 and Ir3 is quite similar to the  case of Ir-$fcc$. This is probably due to the fact, that the Ir2-Ir3 and Ir3-Ir3 coordination, as well as  interatomic distances, are very close to the $fcc$ phase (Ir-Ir distance is $\sim 2.7$~\AA~in the aforementioned cases, see \tref{tab:dist}). 
However, the $n(E_F)$ values on Ir atoms are much lower, comparing to $fcc$ structure ($n(E_F) \simeq$ 6.3 Ry$^{-1}$/spin in Ir-$fcc$), being the effect of enhanced hybridisation near $E_F$.
Noteworthy, Ir1 has a quite different atomic coordination, with respect to Ir2 and Ir3 positions, being surrounded practically only by 4 boron atoms (B1). It is clearly reflected by the apparently different DOS shape below $E_F$.

\begin{figure}[b]
\includegraphics[width=0.7\textwidth]{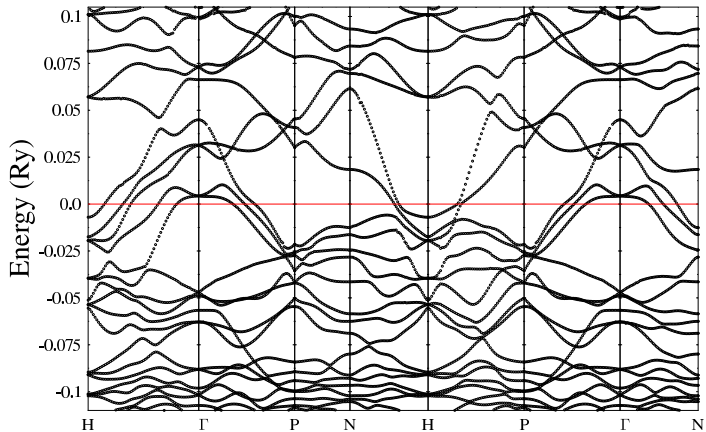}
\caption{Dispersion curves near the $E_F=0$ in Mg$_{10}$Ir$_{19}$B$_{16}$.}\label{fig:bands}
\end{figure}

Electronic dispersion curves near the Fermi level are presented in \fref{fig:bands}. The bands located below $-0.1$~Ry (partly showed) are very flat, and form the narrow large DOS peaks, seen already in \fref{fig:dos}. Conversely, the bands that cross $E_F$ are quite dispersive, which results in the low DOS in this energy range. Noteworthy, there is an energy gap along the P-N direction (parallel to the $k_z$ axis in the reciprocal space). This may suggest some anisotropic transport properties of this compound.

\begin{table}[t]
\caption{\label{tab:dist} The smallest interatomic distances between atoms (in \AA) and atomic positions in Mg$_{10}$Ir$_{19}$B$_{16}$. The Ir-Ir distance in Ir-$fcc$ is about 2.7 \AA, as between Ir2-Ir3 and Ir3-Ir3.}
\lineup
\begin{small}
\begin{tabular}{@{}llllllllcl}
\br
&     Ir1& Ir2 & Ir3 & B1  & B2 & Mg1  &Mg2 & Atomic position &  Site \\
\mr
Ir1  & 9.1&  5.9&  3.9&  2.1&  5.0&  3.1&  3.6& (0, 0, 0)                  & \0(2a)  \\    
Ir2  &    &  3.8&  2.7&  4.5&  2.2&  4.0&  3.1& (0, 0.25, 0.5)             & (12d)   \\    
Ir3  &    &     &  2.7&  2.1&  2.2&  4.0&  2.9& (0.0702, 0.2525, 0.2525)   & (24g)   \\    
B1   &    &     &     &  3.4&  3.3&  3.0&  3.0& (0.3331, 0.3331, 0.3331)   & \0(8c)  \\    
B2   &    &     &     &     &  3.3&  2.6&  2.5& (0, 0, 0.3473)             & (24g)   \\    
Mg1  &    &     &     &     &     &  5.0&  3.1& (0.1127, 0.1127, 0.1127)   & \0(8c)  \\    
Mg2  &    &     &     &     &     &     &  3.1& (0.1639, 0.1639, 0.4140)   & (12e)   \\    
\br
\end{tabular}
\end{small}
\end{table}

The calculated McMillan-Hopfield factors for all sites, with contributions from each scattering channel ($l\rightarrow l+1$), are presented in \tref{tab:eta}.
Estimation of the electron-phonon coupling constant $\lambda$, using MH parameters, requires also the knowledge of average phonon frequencies. For such a large structure, containing 90 atoms in the cubic unit cell, phonon spectra calculations are difficult to be carried out. Nevertheless we can try to investigate the strength of electron-phonon interaction assuming reasonable values of $\langle \omega_i^2 \rangle$ and studying 
their influence on estimated $\lambda$ values. 
Debye frequencies $\Theta$ of monoatomic crystals of iridium ($fcc$), boron (rhomboedral) and magnesium ($hcp$) may be helpful for choosing sensible phonon frequency range for our discussion.
At first we assume, that the same type of atoms, occupying different sites in the cell, have similar average vibrational frequencies. Iridium, as the heaviest element, is expected to have the lowest $\langle \omega_i^2 \rangle$, Debye frequency of metallic iridium is about $\Theta_{Ir} \simeq 36$~meV (420~K)~\cite{deb-ir}. In contrast, the lightest boron is certainly expected to have the highest phonon frequencies, and for crystalline boron $\Theta_{B} \simeq 100$~meV (1200~K)~\cite{deb-b}.
Finally, magnesium average frequencies are expected to locate between the values of iridium and boron. Debye frequency of crystalline Mg is rather low: $\Theta_{Mg} \simeq 34$~meV (400~K)~\cite{deb-ir}.
We may also recall, that average square phonon frequency is often estimated as $\langle \omega^2 \rangle \simeq {1\over2} \Theta^2$, which is a good approximation in monoatomic structures. In our case, these values may also be helpful for choosing feasible range of $\langle \omega_i^2 \rangle$.

\begin{figure}[t]
\includegraphics[width=\textwidth]{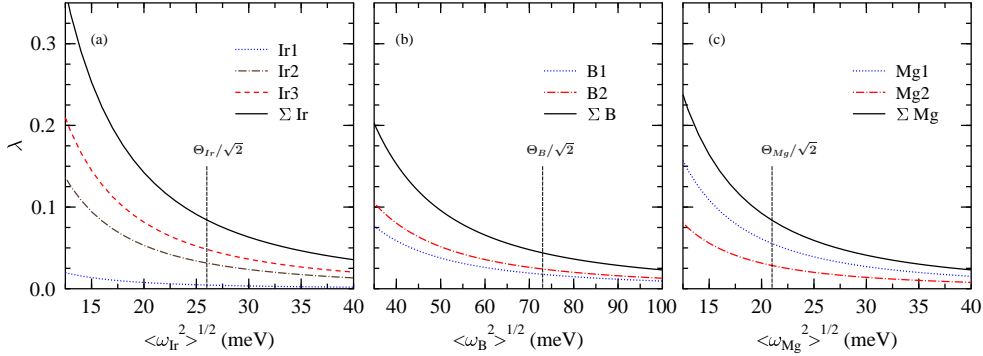}
\caption{Contributions to the electron-phonon coupling constant $\lambda$ in Mg$_{10}$Ir$_{19}$B$_{16}$ from iridium (a), boron (b) and magnesium (c) as a function of a root of average square frequency. Top curve on each graph is a sum of contributions from particular sites. Vertical lines mark the value of $\Theta_i/\sqrt{2}$. }\label{fig:lambda}
\end{figure}

\Fref{fig:lambda} presents estimated electron-phonon coupling constant $\lambda$, associated with particular crystal sites, plotted as a function of average square phonon frequency. For each type of atoms, a wide frequency range was chosen, to illustrate the changeability of $\lambda$. In the case of iridium, the largest contribution to $\lambda$ comes from Ir3 sublattice, due to the large population of this site.
Among boron and magnesium sites, the B2 and Mg1 atoms provide the largest contributions. However, one has to remember that this comparison is valid only if we assume identical frequencies for the same atoms at different sites. In this simplified analysis, we may also plot the overall coupling constant for constituent atoms, by adding the contributions from each site, which is illustrated by solid lines in \fref{fig:lambda}.

\begin{figure}[t]
\includegraphics[width=0.5\textwidth]{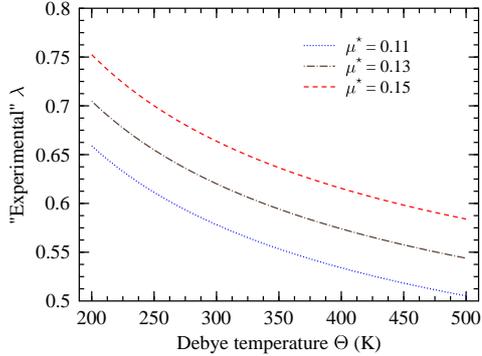}
\caption{"Experimental" value of the electron-phonon coupling constant $\lambda$, evaluated from the McMillan formula, using observed $T_C = 4.5$~K, plotted as a function of Debye temperature $\Theta$ for three values of Coulomb pseudopotential parameters $\mu^{\star}$.}\label{fig:lambda2}
\end{figure}

As we can see, the obtained partial coupling constants are not high for each atom.
At the moment, we are not aware of any experimental findings of the EPC constant in this compound. However, we can try to estimate the range of "experimental" $\lambda$, analysing the magnitude of the observed critical temperature. If we assume, that we are dealing with BCS-type superconductivity, we may substitute the experimental value of $T_C = 4.5$~K  into the McMillan formula for $T_C$~\cite{mcm}:
\begin{equation}\label{eq:tc}
T_C =  \frac{\Theta}{1.45}\exp\left[-\frac{1.04(1+\lambda)}{\lambda-\mu^{\star}(1+0.62\lambda)}\right].
\end{equation}
However, because the Debye temperature $\Theta$ of Mg$_{10}$Ir$_{19}$B$_{16}$ is not known yet, we plot $\lambda$ in \fref{fig:lambda2} as a function of $\Theta$, for typical values of Coulomb pseudopotential parameter $\mu^{\star}$.
The resulting EPC constant  $\lambda$ varies between 0.5 and 0.75,
and for e.g. $\Theta = 250$~K we obtain $\lambda \simeq 0.60$, whereas for $\Theta = 350$~K we get $\lambda \simeq 0.55$ (for $\mu^{\star} = 0.11$), as one can see in the \fref{fig:lambda2}.
Thus, if Mg$_{10}$Ir$_{19}$B$_{16}$ is treated as a conventional superconductor, the lower limit of $\lambda$ is expected to be 0.5.
In order to get this value from our RMTA calculations, relatively low average frequencies for all constituent atoms are required. If we take the following $\sqrt{\langle \omega_i^2 \rangle}$: \makebox{14 meV for Ir}, \makebox{50 meV for B} and \makebox{20 meV for Mg}, we obtain $\lambda_{Ir} \simeq 0.3$, $\lambda_{B} \simeq 0.1$ and $\lambda_{Mg} \simeq 0.1$, which gives the expected lower limit of EPC constant ($\lambda \simeq 0.5$).
If higher values of $\lambda$ are experimentally observed, it will indicate either additional contributions, neglected within the RMTA framework, or even lower phonon frequencies, than the values used in the aforementioned estimations. Especially iridium contribution is sensitive to the change of $\sqrt{\langle \omega_{Ir}^2 \rangle}$ parameter, i.e. increasing it to 25 meV (probably an upper limit) results in decrease of $\lambda_{Ir}$ to 0.1, that is to the value found for other sublattices. 
Generally, our calculations indicates, that iridium sublattices seem to be the most important for the onset of superconductivity in Mg$_{10}$Ir$_{19}$B$_{16}$.

Finally, we shortly comment the possible influence of the lack of inversion symmetry and spin-orbit (SO) interaction on the electronic structure and superconductivity.
The absence of a centre of inversion in a superconductor affects the symmetry of  superconducting state, allowing for an admixture of singlet and triplet components~\cite{sym1, sym2}.
Because the singlet pairing is based on the time-reversal symmetry~\cite{anderson59}, which is present as long as the compound is not magnetic, mainly the triplet channel is affected by the lack of inversion, and the superconductivity may be even suppressed, see e.g.~\cite{anderson,sym1,sym2}.
The SO interaction, which mixes the initial spin-up and spin-down electronic states,
plays an additional role. It was found, that it may control the mixing of parity of the superconducting state~\cite{sym2}.
This seems to be the case of Li$_2$Pd$_3$B and Li$_2$Pt$_3$B compounds, where specific heat~\cite{lipdb-spheat} and NMR~\cite{lipdb-nmr} measurements strongly support phonon-mediated isotropic superconductivity, while penetration depth measurements suggests an admixture of spin-singlet and triplet components in the superconducting energy gap~\cite{lipdb-penetr}, with larger triplet component in the Pt-case.
This kind of experimental study for Mg$_{10}$Ir$_{19}$B$_{16}$ should be prior to the theoretical discussion of gap symmetry in this compound.

As far as the band structure of Mg$_{10}$Ir$_{19}$B$_{16}$ is concerned, the modifications due to the SO interaction are not expected to significantly affect the obtained values of DOS and MH parameters.
For metallic $fcc$ iridium, our value of MH parameter, calculated also neglecting SO interaction, $\eta_{Ir} \simeq 135$~mRy/(a.u.)$^2$, gives the correct magnitude of the electron-phonon coupling constant $\lambda = 0.32$, comparing to the observed $\lambda = 0.34$~\cite{mcm}. Here, the average square frequency was estimated from the formula $\langle \omega^2 \rangle \simeq {1\over2} \Theta^2$, using the Debye temperature $\Theta = 420$~K.

\section{Summary and conclusions}
The results of LDA electronic structure calculations of new Mg$_{10}$Ir$_{19}$B$_{16}$ superconductor were presented. The main contributions to densities of states near $E_F$ are provided by iridium atoms. The electron-phonon coupling constant $\lambda$ was roughly estimated, using the calculated McMillan-Hopfield parameters and qualitative discussion of average phonon frequencies. 
We discussed the relation of the experimental transition temperature and the magnitude of EPC coupling.
Within the rigid-muffin-tin approximation, the main contribution to $\lambda$ comes from iridium, with smaller contributions from boron and magnesium. If any information about the dynamic properties of atoms in this compound become available, our analysis presented of \fref{fig:lambda} will allow to find better theoretical estimation of $\lambda$. The location of $E_F$ on the slope of Ir DOS peak leads to rough suggestion, that hole doping on iridium sites, e.g. with rhodium, may increase the densities and MH parameters.

\ack
We would like to thank dr Tomasz Klimczuk for helpful discussions. This work was partly supported by the Polish Ministry of Science and Higher Education (PhD grant).

\end{document}